\documentclass[conference]{IEEEtran}
\IEEEoverridecommandlockouts
\usepackage{cite}
\usepackage{amsmath,amssymb,amsfonts}
\usepackage{algorithmic}
\usepackage{diagbox}
\usepackage{graphicx}
\usepackage{subcaption}
\usepackage{textcomp}
\usepackage{xcolor}
\usepackage{pifont} 
\usepackage{booktabs}
\usepackage{mathrsfs}  
\usepackage{array}  
\def\BibTeX{{\rm B\kern-.05em{\sc i\kern-.025em b}\kern-.08em
    T\kern-.1667em\lower.7ex\hbox{E}\kern-.125emX}}
\begin{document}

\title{Automatic State Machine Inference for Binary Protocol Reverse Engineering\\

\thanks{This work is supported by the National Key Research and Development Program of China (No.2023YFB3107605), the National Natural Science Foundation of China (No.U23B2024), the Pandeng Project of IIE CAS and the Youth Innovation Promotion Association CAS (No.2021154)


}
}
\author{
\IEEEauthorblockN{
Junhai Yang\IEEEauthorrefmark{1}\IEEEauthorrefmark{2}\IEEEauthorrefmark{3}, 
Fenghua Li\IEEEauthorrefmark{1}\IEEEauthorrefmark{2}\IEEEauthorrefmark{3}, 
Yixuan Zhang\IEEEauthorrefmark{1}\IEEEauthorrefmark{2}\IEEEauthorrefmark{3}, 
Junhao Zhang\IEEEauthorrefmark{1}\IEEEauthorrefmark{2}\IEEEauthorrefmark{3}, 
Liang Fang\IEEEauthorrefmark{1}\IEEEauthorrefmark{2}\IEEEauthorrefmark{3}\IEEEauthorrefmark{4}, 
Yunchuan Guo\IEEEauthorrefmark{1}\IEEEauthorrefmark{2}\IEEEauthorrefmark{3}} 
\IEEEauthorblockA{\IEEEauthorrefmark{1}Institute of Information Engineering, Chinese Academy of Sciences, Beijing, China}
\IEEEauthorblockA{\IEEEauthorrefmark{2}School of Cyber Security, University of Chinese Academy of Sciences, Beijing, China}
        \IEEEauthorblockA{\IEEEauthorrefmark{3}Key Laboratory of Cyberspace Security Defense, Beijing, China}
        \IEEEauthorblockA{\IEEEauthorrefmark{4}Corresponding Author}
        Email: \emph{\{yangjunhai, lifenghua, zhangyixuan, zhangjunhao, fangliang, guoyunchuan\}@iie.ac.cn}
}

\maketitle

\begin{abstract}
Protocol Reverse Engineering (PRE) is used to analyze the protocol by inferring the protocol structure and behavior. However, current PRE methods mainly focus on field identification within a single protocol and neglect Protocol State Machine (PSM) analysis in a mixed protocols environment. This will lead to insufficient analysis of protocols’ abnormal behavior and potential vulnerabilities, which are crucial for detecting and defending against new attack patterns. To address these challenges, we propose an automatic PSM inference framework for unknown protocols, including a fuzzy membership-based auto-converging DBSCAN algorithm for protocol format clustering, followed by a session clustering algorithm based on Needleman-Wunsch and K-Medoids algorithms to classify sessions by protocol type. Finally, we refined probabilistic PSM algorithm to infer the protocol states and the transition conditions between these states. Experimental results show that compared with existing PRE techniques, our method can infer PSMs while enables more precise classification of protocols.
\end{abstract}

\begin{IEEEkeywords}
protocol reverse engineering, state machine inference, protocol classification
\end{IEEEkeywords}

\begin{figure*}[ht]
    \centering
    \includegraphics[width=\textwidth]{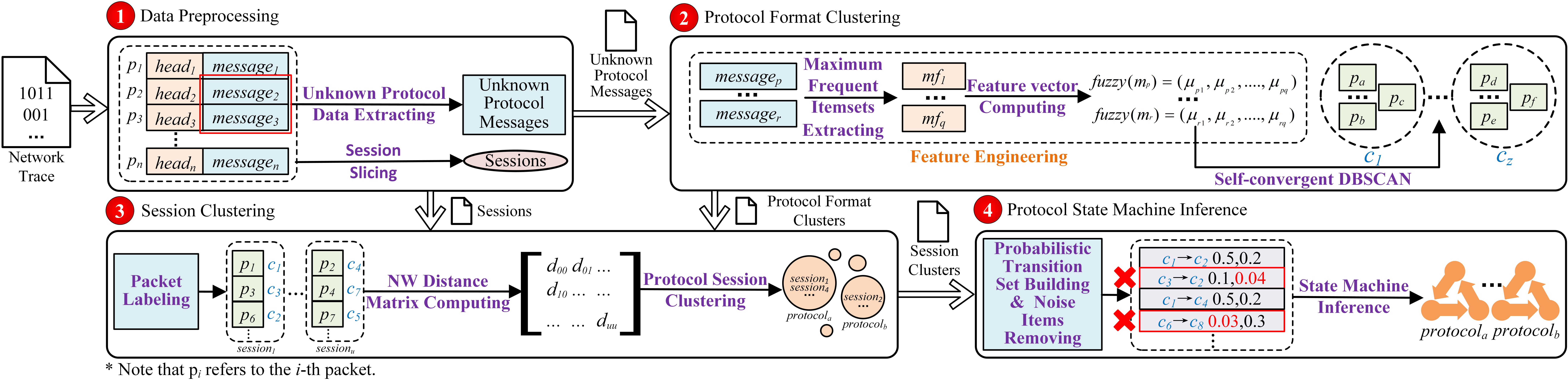} 
    \caption{System architecture of automatic PSM inference framework for unknown protocols} 
    \label{fig.1} 
\end{figure*}

\section{Introduction}
With the rapid development of network applications, proprietary protocols have been widely used, leading to increased complexity and diversity in public network traffic. According to 2023 reports from Cloudflare\footnote{https://radar.cloudflare.com/year-in-review/2023} and Palo Alto Networks\footnote{https://www.paloaltonetworks.com/resources/research/unit-42-attack-surface-threat-report-2023}, about 30\% of global traffic is composed of unknown 
proprietary protocol traffic. This presents significant challenges for malware behavior analysis\cite{malware1,malware2,malware3}, intrusion detection\cite{inrusion1},\cite{inrusion2}, vulnerability scanning\cite{-vulnerabilityScanning,Delta-vulnerabilityScanning}, fuzzing\cite{fuzzing} and automated exploit generation\cite{automatic1,automatic2,automated3}. 

\textbf{Protocol Reverse Engineering (PRE)} aims to analyze the structure, functionality and behavior of protocols, revealing potential security threats and patterns in data transmission. Existing PRE methods fall into two categories\cite{Netplier}: program-based\cite{DYNPRE,program1,program2} and network trace-based\cite{BinaryInferno,Discoverer,Nemesys-networkTrace,Netplier,Netzob} methods. Program-based methods require interaction with protocol binary programs, but these are often inaccessible due to protective measures implemented by owners, such as encryption or obfuscation. Whereas network trace-based methods rely only on captured traffic, offering a more practical approach.

However, existing network trace-based PRE methods primarily focus on field identification while \textbf{do not give much attention to the analysis of \textbf{Protocol State Machine(PSM)}}. PSM represents the states and transitions in protocol interactions, offering valuable insight into the communication patterns between malware and its servers. This comprehensive understanding of protocol functionality aids in vulnerability detection and the optimization of scanning strategies. Furthermore, it enables the detection of new types of attacks and the generation of attack signatures. Besides, existing PRE solutions \textbf{can only analyze network trace with a single protocol}, whereas current network traffic often consists of multiple unknown protocols.

To address these issues, we proposed an automatic PSM inference framework for unknown protocols from network trace. Our main contributions are as follows:

\begin{itemize}
\item According to cluster analysis and probabilistic PSM, we proposed an automatic PSM inference framework for unknown protocols. First, unknown protocol messages are clustered by protocol format. Then, session clustering is performed to group sessions by protocol type, facilitating PSM inference for mixed protocols. Last, we infer PSMs for each protocol based on session clustering results.


\item We proposed a fuzzy membership-based auto-converging DBSCAN algorithm for protocol format clustering, which extracts feature vectors from protocol messages and automatically converges to the optimal result. And we proposed a session clustering method using the Needleman-Wunsch(NW)\cite{NW} and K-Medoids\cite{K-mer} algorithms to group sessions by protocol type. We enhanced Veritas\cite{veritas} by incorporating total probability, resolving the problem where transitions originating from the same state with only misclassified transitions but could not be identified, significantly improving the accuracy of PSM inference.

\item We tested our system using real network traffic captured from our laboratory gateway and achieved a State Matching Coefficient and Transition Matching Coefficient of 0.96 and 0.93 in PSM inference, respectively.
\end{itemize}

\section{Method}
\subsection{Overview}
Our method can perform PRE on network trace composed of mixed unknown protocols, as illustrated in Fig.~\ref{fig.1}. The entire process is composed of four key steps:

\textbf{Step 1: Data Preprocessing.} We parse known protocol data, extract unknown protocol messages for \textbf{step 2} and slice sessions for \textbf{step 3}.

\textbf{Step 2: Protocol Format Clustering.} We use 
the fast Apriori algorithm\cite{Apriori} to extract maximum frequent itemset and calculate fuzzy membership feature vectors. We then propose an \textbf{auto-converging DBSCAN algorithm (ACDA)} to cluster the feature vectors of unknown protocol messages, classifying them by protocol format.

\textbf{Step 3: Session Clustering.} We label every packet in a session with the clustering results in \textbf{step 2}, obtaining sessions represented by the sequence of protocol format cluster numbers. Then, we utilize the NW and K-Medoids algorithm to cluster the sessions by protocol type.

\textbf{Step 4: Protocol State Machine Inference.} We construct a probabilistic protocol format transition set, filter out low-probability transitions. And converting the protocol format transition set into a state transition set that  describes the interaction process between the protocol client and server. Finally, we automatically infer PSMs in different protocols.

\subsection{Data Preprocessing}
Data preprocessing is conducted based on model matching, utilizing JSON descriptions of various known protocol formats, key field functionalities and PSMs. Through model matching, all known protocol data are identified and labeled, and the others are defined as unknown protocol messages, denoted as $\mathcal{M}=\{m_1,m_2,\dots,m_l\}$, where $l$ represents the number of messages. Next procedure is session slicing. All sessions containing $\mathcal{M}$ are preserved, denoted as $\mathcal{S}=\{s_1,s_2,\dots,s_u\}$, where $u$ is the number of sessions.

\subsection{Protocol Format Clustering}\label{sec:2.3}


First, we use the fast Apriori algorithm to extract maximum frequent itemset, defined as $MFI=\{mf_1,mf_2,\ldots,mf_q\}$, where $q$ is the frequent item count. Items are extracted if they exceed the minimum support, denoted as $ms$, defined in \eqref{eq1}. If items in $MFI$ exhibit containment relationships, the contained items are eliminated. And each item must be contiguous, i.e., a substring of $\mathcal{M}$. Moreover, we only extract frequent items of lengths 1,2,4,8 bytes, because fields are often represented by integers or floating-point numbers in the protocol design.

\begin{equation}
ms = \frac{\left|\{mf\subseteq m_i |  m_i\in \mathcal{M}\}\right |}{\left|\mathcal{M}\right | } \label{eq1}
\end{equation}

Then, we calculate feature vectors for $\mathcal{M}$, denoted as $\mathcal{V}=\{v_1,v_2,\dots,v_l\}$, where $v_i$ is the feature vector of $m_i$ and $v_i=\{\mu_{i1},\mu_{i2},\ldots,\mu_{iq}\}$, where $\mu_{ij}$ is the fuzzy membership function, represents the membership of $mf_j$ to $m_i$, ranging from 0 to 1 and is calculated in \eqref{eq2}.

\begin{equation}
  \mu_{ij} = \frac{LCSS(mf_j,m_i) }{length(mf_j) }  \label{eq2}
\end{equation}
where $LCSS( f_j,m_i)$ is the length of the longest common substring between $mf_j$ and $m_i$.


\begin{equation}
  d(v_i,v_j) = \sqrt{ {\textstyle \sum_{k = 1}^{q}}(\mu_{ik} - \mu_{jk})^2 } \label{eq3}
\end{equation}

Last, we propose ACDA to cluster $\mathcal{V}$ of $\mathcal{M}$ using Euclidean Distance\cite{Oujilide} to measure the difference between feature vectors, as shown in \eqref{eq3}. A convergence function based on Silhouette Coefficient \cite{SC}, denoted as $SC$, evaluates clustering quality. The $eps$ and $minPts$ in DBSCAN are adjusted in each iteration by their step sizes ($eps_s$ and $minPts_s$). And ACDA dynamically adjusts $eps_s$ and $minPts_s$ by the convergence threshold, denoted as $tol$, until $SC$ converges to the optimal value. We set the initial range for $eps$ and $minPts$, along with $eps_s$ and $minPts_s$ and the value of $tol$. During each clustering iteration, improvement is defined as $imp=SC_b-SC_{pb}$, where $SC_b$ is the best $SC$ in the current iteration, and $SC_{pb}$ is the optimal $SC$ from the previous. The dynamic adjustment process of $eps_s$ and $minPts_s$ is shown in \eqref{eq4} and \eqref{eq5}.


\begin{equation}
  eps_s = \begin{cases}
  \text{min}(eps_s \times (1+imp),\alpha) \text{, if } imp>tol \\
  \text{max}(eps_s \times (1-imp),\beta)\text{, if } imp \le tol
\end{cases} \label{eq4}
\end{equation}

\begin{equation}
\resizebox{\columnwidth}{!}{$
  minPts_s = \begin{cases}
  \text{min}(minPts_s \times (1+imp),\gamma), \text{if } imp>tol \\
  \text{max}(minPts_s \times (1-imp),\lambda), \text{if } imp \le tol
\end{cases}
$}
\label{eq5}
\end{equation}
where $\alpha$ and $\beta$ are the max and min constraints of $eps_s$, $\gamma$ and $\lambda$ are for $minPts_s$ to limit extreme big or small values.

The optimal clustering result obtained by ACDA is referred to as protocol format clusters, denoted as $PFC = \{c_1, c_2, \ldots, c_g\}$, where $c_i$ is a cluster number.

\subsection{Session Clustering}

\begin{figure}[h]
    \centering
    \centerline{\includegraphics[width=\linewidth]{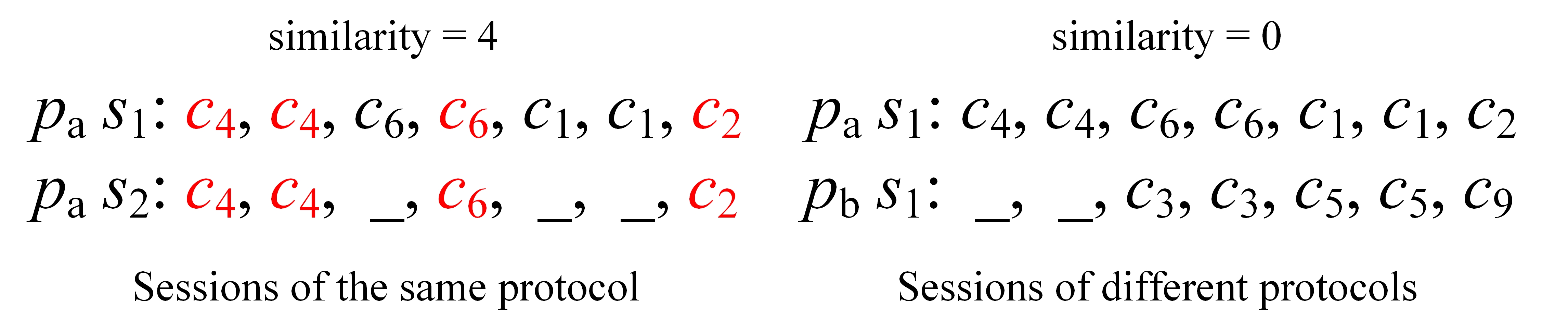}}
    \caption{NW algorithm $PFC$ sequence alignment and similarity calculation.} 
    \label{fig.2} 
\end{figure}

First, sessions from different protocols have distinct protocol formats, while those from the same protocol show high similarity in types and sequence order. So we label packets in $\mathcal{S}$ with $PFC$ numbers, representing each session as a sequence of $PFC$ numbers. Then, despite differences in packet counts or processes between sessions of the same protocol, we use the NW algorithm to align sequences and enhance similarity, with the number of identical $PFC$ numbers at the same positions indicating session similarity, as shown in Fig.~\ref{fig.2}, where $p$ represents protocol and $s$ represents session. The higher the similarity, the closer the two feature vectors are.


Last, we use the K-medoids algorithm for session clustering and use $SC$ to choose the best result. Given that the number of protocol types is smaller than the number of protocol formats, the range of the cluster number, denoted as $cn$, for the K-medoids algorithm is set from 1 to $length(PFC)$.

\subsection{Protocol State Machine Inference}
We construct Protocol Format Transition Set (PFTS) and refine probabilistic PSM from Veritas\cite{veritas} to eliminate noise. Noise in real traffic and misclassification during clustering can lead to inaccurate PSM inference. Despite Veritas's state probability, denoted as $P_s$, we introduce total probability, denoted as $P_t$, resolving the problem where transitions originating from one state consisted entirely of noise transitions but couldn't be identified, enhancing the accuracy and robustness of PSM inference. The definitions of $P_s$ and $P_t$ are calculated in \eqref{eq6}.

\begin{equation}
    P_s(c_i\rightarrow c_j) = \frac{N_{c_i\rightarrow c_j}}{N_{c_i\rightarrow all}} \ \ \     P_t (c_i\rightarrow c_j) = \frac{N_{c_i\rightarrow c_j}}{N_{set}}\label{eq6}
\end{equation}
where $N_{c_i\rightarrow c_j}$ is the number of transitions from $c_i$ to $c_j$, $N_{c_i\rightarrow all}$ is the number of transitions from $c_i$ to other cluster numbers and $N_{set}$ represents the total number of transitions.

We define the thresholds of $P_s$ and $P_t$ as $t_{ps}$ and $t_{pt}$. Transitions where $P_s$ is less than $t_{ps}$ or $P_t$ is less than $t_{pt}$ are considered as noise and removed.

PFTS describes the transitions between protocol formats, but we need to depict the interaction between the protocol client and server in the PSM. Therefore, we convert PFTS into a PSM where the states represent the server and client, and the transitions correspond to protocol formats.


\section{Experiments and Analysis}
\subsection{Environment and Dataset}
We run our experiment on a platform with Intel(R) Core(TM) i7-11800H 2.30GHz CPU, 64GB RAM and CentOS7.2.9 System. The program is developed by Python 3 and C language. We capture 30 seconds of the network traffic data (including DNS, SMTP, POP3,TLSv1.2, HTTP, HTTPS etc.) from our laboratory’s gateway as the experimental dataset. To evaluate the performance and effectiveness of our system on both binary and text protocols, we use TLSv1.2 and SMTP protocol as black-box for evaluation.

The TLSv1.2 dataset contains four protocol formats: Handshake, Change Cipher Spec, Application Data and Alert. The SMTP dataset includes five protocol formats: HELO, MAIL FROM, RCPT TO, DATA and QUIT. The real PSMs of TLSv1.2 and SMTP are shown in Fig.~\ref{fig.7}.

\begin{figure*}[ht]
    \centering
    \includegraphics[width=\textwidth]{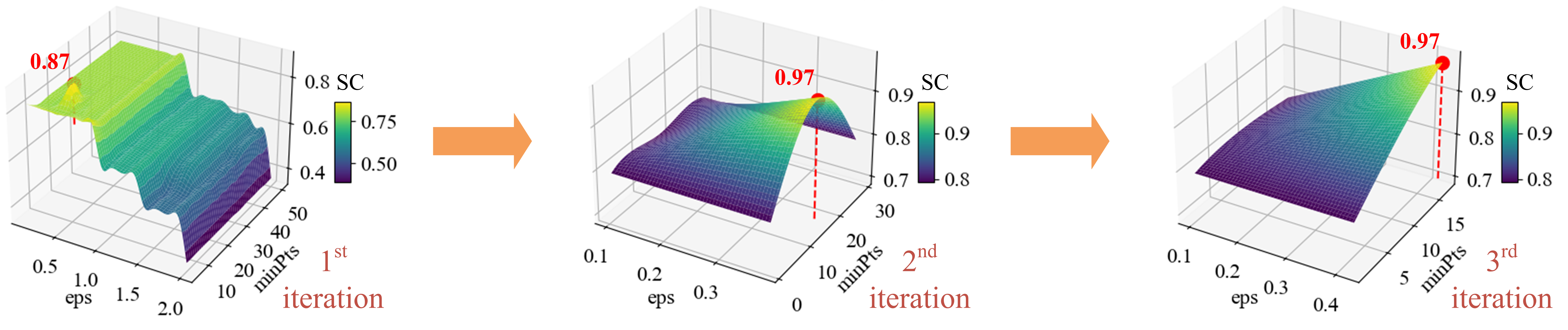} 
    \caption{The optimal clustering result calculation for the ACDA.} 
    \label{fig.3} 
\end{figure*}


\begin{figure}[htbp]
    \centering
    \begin{subfigure}[b]{0.49\linewidth} 
        \centering
        \includegraphics[width=\textwidth]{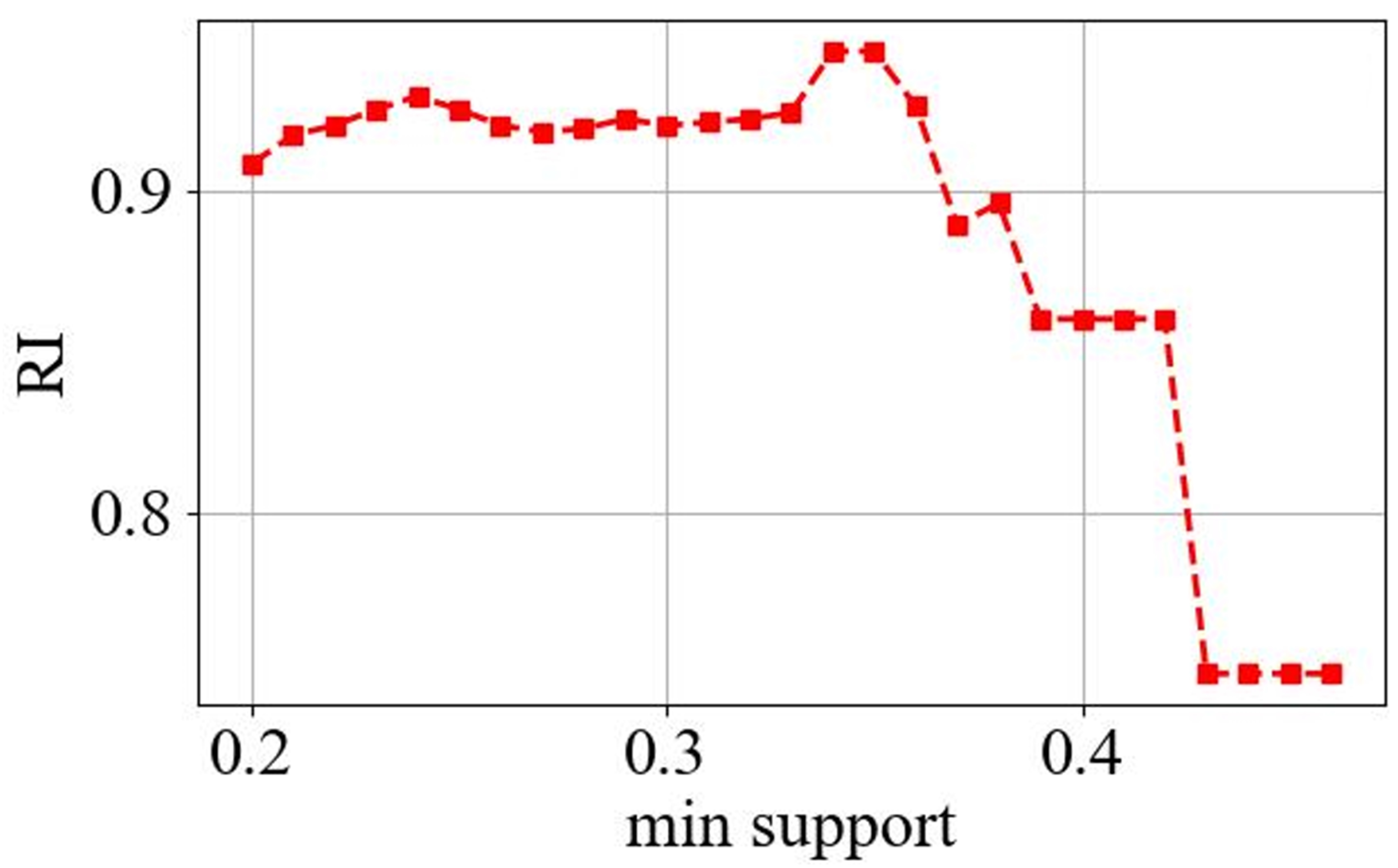} 
        \caption{$min \ support$ Selection}
        \label{fig4:sub1}
    \end{subfigure}%
    \hfill 
    \begin{subfigure}[b]{0.49\linewidth} 
        \centering
        \includegraphics[width=\textwidth]{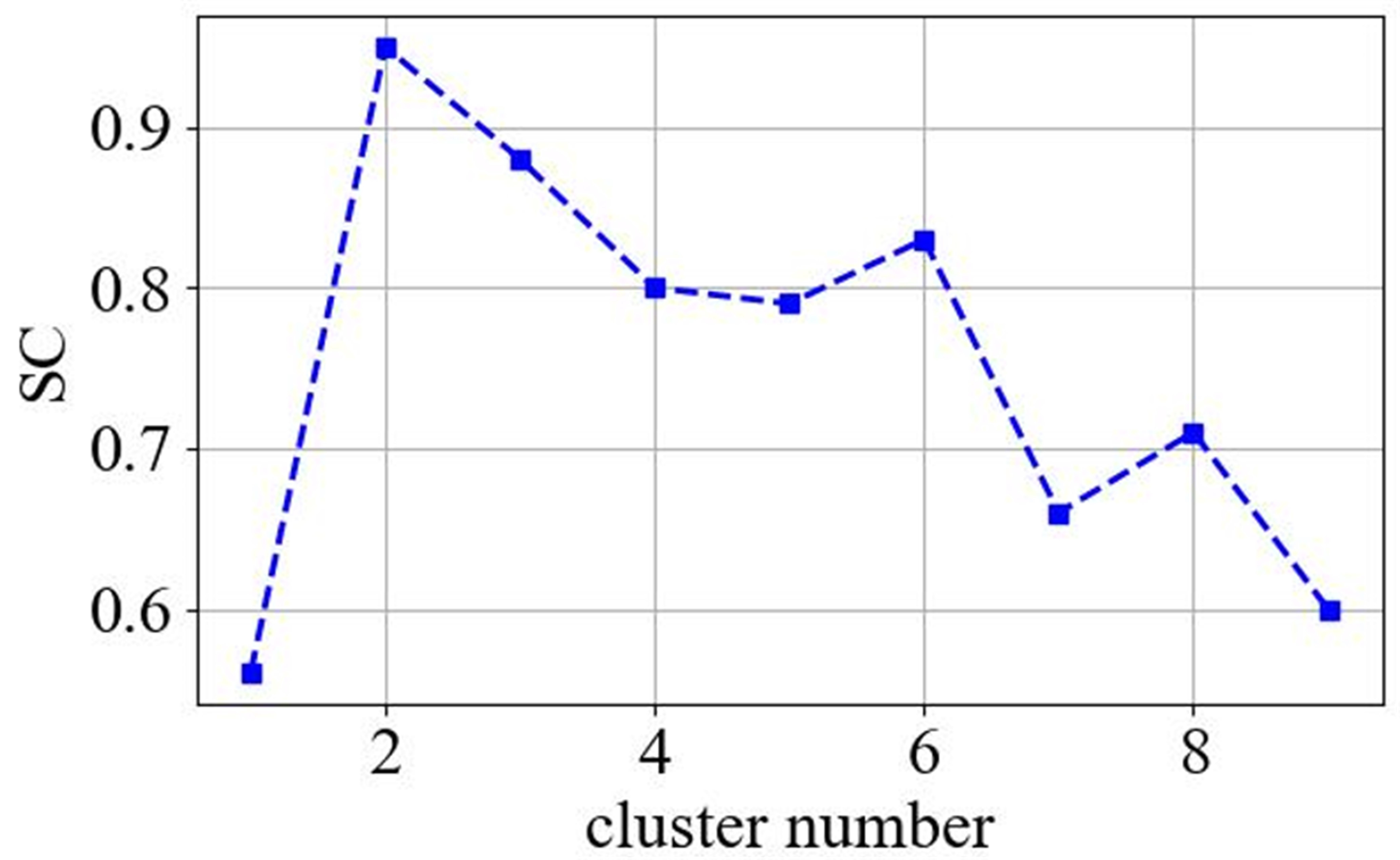}
        \caption{$cluster \ number$ selection}
        \label{fig4:sub2}
    \end{subfigure}
    \caption{The optimal parameters selection}
    \label{fig.4}
\end{figure}

\subsection{Metrics}
\subsubsection{Clustering Evaluation}
To evaluate the performance of protocol format clustering and session clustering, we use two common performance metrics in unsupervised learning: $SC$ and Rand Index\cite{randIndex}, denoted as $RI$. 

$SC$ measures intra-cluster consistency and inter-cluster separation, reflecting how well data points fit within their clusters. The average $SC$, ranging from -1 to 1, indicates clustering quality, with higher values signifying better results. The definition of $SC$ is given in \eqref{eq7}.

\begin{equation}
  SC(i) = \frac{b(i) - a(i)}{\text{max}(a(i), b(i))}    \label{eq7}
\end{equation}
where $a_i$ is the average intra-cluster distance and $b_i$ is the average distance between $c_i$ and the nearest other cluster.



$RI$ quantifies the alignment between clustering results and ground truth labels, ranging from 0 to 1. A value of 1 indicates a perfect match, and 0 indicates complete discordance. The definition of $RI$ is given in \eqref{eq10}.


\begin{equation}
  RI = \frac{TP+TN}{TP+TN+FP+FN}  \label{eq10}
\end{equation}
where $TP$,$TN$,$TP$,$FN$ are parameters in confusion matrix.

\subsubsection{PSM Inference Evaluation}
To assess the similarity between real PSM and inferred PSM, we propose \textbf{State Matching Coefficient (SMC)} and \textbf{Transition Matching Coefficient (TMC)}. They range from 0 to 1, a high matching rate indicates that PSM inference accurately reflects the actual behavior of the protocol. $SMC$ and $TMC$ are defined in\eqref{eq11} and \eqref{eq12}.

\begin{equation}
  SMC(psm_i , psm_j) = \frac{2 \times |S(i) \cap  S(j)|}{N_S(i) + N_S(j)}  \label{eq11}
\end{equation}

\begin{equation}
  TMC(psm_i , psm_j) = \frac{2 \times |T(i) \cap  T(j)|}{N_T(i) + N_T(j)}     \label{eq12}
\end{equation}
where $S(i)$ is the set of states in $psm_i$, $N_S(i)$ is the number of states in $psm_i$. $T(i)$ is the set of transitions in $psm_i$ and $N_T(i)$ is the number of transitions in $psm_i$. 


\subsection{Functionality Comparison}
We compared our system with state-of-the-art PRE methods, including BinaryInferno, Netzob, Netplier and Discoverer. The results, shown in Table \eqref{tab1}, demonstrate that our approach outperforms these methods.

\begin{table}[htbp]
\centering
\caption{Functionality comparison.}
\begin{tabular}{|>{\centering\arraybackslash}m{0.39\linewidth}|>{\centering\arraybackslash}m{0.06\linewidth}|>{\centering\arraybackslash}m{0.05\linewidth}|>{\centering\arraybackslash}m{0.05\linewidth}|>{\centering\arraybackslash}m{0.05\linewidth}|>{\centering\arraybackslash}m{0.05\linewidth}|}
\hline
\diagbox{\textbf{Functions}}{\textbf{Methods}}
&\textbf{Ours}&Ref.\cite{BinaryInferno}&Ref.\cite{Netzob}&Ref.\cite{Netplier}&Ref.\cite{Discoverer}\\
\hline

No Need of Keyword Support & \ding{51} & \ding{51} & \ding{51} & \ding{55} & \ding{51} \\ \hline
No Need of Delimiter Input & \ding{51} & \ding{51} & \ding{55} & \ding{51} & \ding{51} \\ \hline
Protocol Format Clustering & \ding{51} & \ding{55} & \ding{51} & \ding{55} & \ding{51} \\ \hline
Session Clustering & \ding{51} & \ding{55} & \ding{55} & \ding{55} & \ding{55} \\ \hline

State Machine Inference & \ding{51} & \ding{55} & \ding{55} & \ding{51} & \ding{51} \\ \hline

Mixed Protocols Support & \ding{51} & \ding{55} & \ding{55} & \ding{55} & \ding{55} \\ \hline

\end{tabular}
\label{tab1}
\footnotesize{*Note: BinaryInferno\cite{BinaryInferno}, Netzob\cite{Netzob}, Netplier\cite{Netplier}, Discoverer\cite{Discoverer}}
\end{table}

\subsection{Experiment Results}
\subsubsection{Protocol Format Clustering}\label{section PFC}
We set $eps$ from 0.1 to 2.0, $minPts$ from 5 to 50, $eps_s$ is 0.1, $minPts_s$ is 5, and $tol$ is 0.01 for ACDA. For the fast Apriori algorithm, the initial range for $ms$ is 0.2-0.45, with a time complexity $O(q \times 2^{cn})$, where $q$ is the dataset size, and $cn$ is the candidate set size. Too low $ms$ produces many redundant features, which worsens the clustering results and wastes time. If $ms$ is too high, $MFI$ becomes too small, making it hard to find useful features. 

We use labeled TLS1.2 dataset for protocol format clustering to choose the best $ms$. The best result is achieved with a $ms$ value of 0.35, where the $RI$ is highest at 0.94, as shown in Fig.~\ref{fig4:sub1}. Therefore, we select 0.35 as $ms$'s value.

We use a mixed dataset of TLS1.2 and SMTP in protocol format clustering. The results are shown in Fig.~\ref{fig.3}. ADCA performs three iterations. The $2^{nd}$ and $3^{rd}$ iterations produce the same best $SC$, with $eps$ = 0.42 and $minPts$ = 17. Comparing with the true labels, we obtain $RI$ of 0.94. The clustering generates 9 clusters, which matches the true labels.

\subsubsection{Session Clustering}
113 sessions are generated in \textbf{step 1}. We label the packets in all sessions based on the clustering results from \textbf{step 2}. We cluster 9 $PFC$ in \eqref{section PFC}. So we set $cn$ from 1 to 9, the results are shown in Fig.~\ref{fig4:sub2}. We find that when k = 2, $SC$ reaches the highest value of 0.95, indicating the best result. Based on this result, we divide all sessions into two categories, marked as $protocol_a$ and $protocol_b$.



\begin{figure}[htbp]
    \centering
    \centerline{\includegraphics[width=\linewidth]{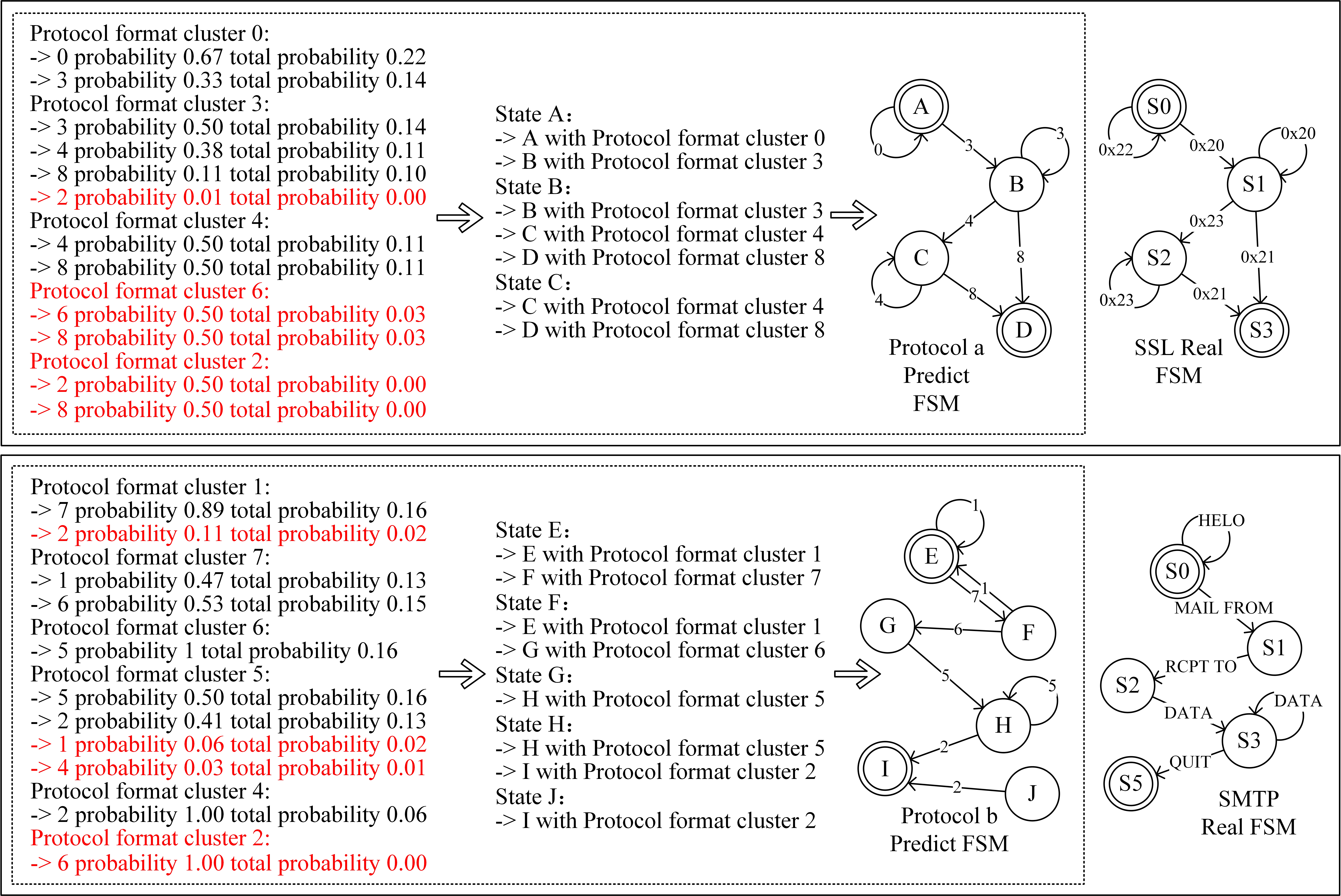}}
    \caption{The process of PSM inference.} 
    \label{fig.7} 
\end{figure}

\subsubsection{Protocol State Machine Inference}
First, we set $t_{ps}$ = 0.05 and $t_{pt}$ = 0.05 to extract probability PFTS and perform noise removal.

To assess the accuracy of the inferred PSMs, we compare them with the actual PSMs of TLSv1.2 and SMTP. The results show that the inferred $protocol_a$ PSM closely matches TLSv1.2, while the inferred $protocol_b$ PSM matches SMTP. TLSv1.2 and $protocol_a$ has an $SMC$ of 1.0 and a $TMC$ of 1.0; SMTP and $protocol_b$ has an $SMC$ of 0.91 and a $TMC$ of 0.86. The total process is shown in Fig.~\ref{fig.7}.


\section{Conclusion}
We proposed an automatic PSM inference framework for network trace. Our method achieved precise PSM inference for mixed unknown protocols and is suitable for both binary and text protocols, outperforming existing PRE methods.

\nocite{*}
\bibliographystyle{IEEEtran}
\bibliography{IEEEabrv,reference}

\end{document}